\documentclass[12pt,a4paper]{article}

\usepackage{lineno}

\usepackage{graphicx}
\usepackage{xspace}
\usepackage{color}
\usepackage{colortbl}
\usepackage{amsmath}
\usepackage{url}
\usepackage{ifthen}

\newboolean{pdflatex}
\setboolean{pdflatex}{false}

\newboolean{uprightparticles}
\setboolean{uprightparticles}{false}
\usepackage{amssymb}
\usepackage{amsfonts}
\usepackage{upgreek}
\usepackage{hepunits}

\usepackage{hyperref}
\hypersetup{
    breaklinks=true,
    plainpages=false,
    bookmarksnumbered = true,
    pdfpagemode       = UseNone,
    pdfstartview      = FitH,
    pdfpagelayout     = SinglePage,
    colorlinks        = true,
    linkcolor         = blue,
    urlcolor          = red,
    citecolor = red
    }
   
\def\lhcb {LHCb\xspace}
\def\ux85 {UX85\xspace}

\def\lhc {LHC\xspace}

\def\babar  {BABAR\xspace}

\def\rich   {RICH\xspace}

\def\lone   {L0\xspace}

\ifthenelse{\boolean{uprightparticles}}%
{

 \def\Ppi         {\ensuremath{\uppi}\xspace}                 
                  
 \def\Prho        {\ensuremath{\uprho}\xspace}

 \def\PDelta      {\ensuremath{\Delta}\xspace}                 
 \def\PXi      {\ensuremath{\Xi}\xspace}                 
 \def\PLambda      {\ensuremath{\Lambda}\xspace}                 
 \def\PSigma      {\ensuremath{\Sigma}\xspace}                 
 \def\POmega      {\ensuremath{\Omega}\xspace}                 
 \def\PUpsilon      {\ensuremath{\Upsilon}\xspace}                 
 
 \def\PB      {\ensuremath{\mathrm{B}}\xspace}                 
                  
 \def\PD      {\ensuremath{\mathrm{D}}\xspace}

 \def\PK      {\ensuremath{\mathrm{K}}\xspace}

 \def\PV      {\ensuremath{\mathrm{V}}\xspace}                 
                  
 \def\PX      {\ensuremath{\mathrm{X}}\xspace}

 \def\Pb      {\ensuremath{\mathrm{b}}\xspace}                 
 \def\Pc      {\ensuremath{\mathrm{c}}\xspace}

 \def\Pi      {\ensuremath{\mathrm{i}}\xspace}

 \def\Pp      {\ensuremath{\mathrm{p}}\xspace}

 \def\Pu      {\ensuremath{\mathrm{u}}\xspace}

}
{

 \def\Ppi         {\ensuremath{\pi}\xspace}                 
                  
 \def\Prho        {\ensuremath{\rho}\xspace}

 \mathchardef\PDelta="7101
 \mathchardef\PXi="7104
 \mathchardef\PLambda="7103
 \mathchardef\PSigma="7106
 \mathchardef\POmega="710A
 \mathchardef\PUpsilon="7107
                  
 \def\PB      {\ensuremath{B}\xspace}                 
                  
 \def\PD      {\ensuremath{D}\xspace}

 \def\PK      {\ensuremath{K}\xspace}

 \def\PV      {\ensuremath{V}\xspace}                 
                  
 \def\PX      {\ensuremath{X}\xspace}

 \def\Pb      {\ensuremath{b}\xspace}                 
 \def\Pc      {\ensuremath{c}\xspace}

 \def\Pi      {\ensuremath{i}\xspace}

 \def\Pp      {\ensuremath{p}\xspace}

 \def\Pu      {\ensuremath{u}\xspace}

}

\def\u     {\ensuremath{\Pu}\xspace}

\def\c     {\ensuremath{\Pc}\xspace}

\def\b     {\ensuremath{\Pb}\xspace}

\def\pion  {\ensuremath{\Ppi}\xspace}

\def\pip   {\ensuremath{\pion^+}\xspace}
\def\pim   {\ensuremath{\pion^-}\xspace}
\def\pipi  {\ensuremath{\pion^+\pion^-}\xspace}

\def\kaon  {\ensuremath{\PK}\xspace}
  \def\Kbar  {\kern 0.2em\overline{\kern -0.2em \PK}{}\xspace}

\def\Kz    {\ensuremath{\kaon^0}\xspace}
\def\Kzb   {\ensuremath{\Kbar^0}\xspace}
\def\KzKzb {\ensuremath{\Kz \kern -0.16em \Kzb}\xspace}
\def\Kp    {\ensuremath{\kaon^+}\xspace}
\def\Km    {\ensuremath{\kaon^-}\xspace}

\def\KpKm  {\ensuremath{\Kp \kern -0.16em \Km}\xspace}
\def\KS    {\ensuremath{\kaon^0_{\rm\scriptscriptstyle S}}\xspace} 
 
\def\Kstarz  {\ensuremath{\kaon^{*0}}\xspace}
\def\Kstarzb {\ensuremath{\Kbar^{*0}}\xspace}

  \def\Dbar    {\kern 0.2em\overline{\kern -0.2em \PD}{}\xspace}
\def\D       {\ensuremath{\PD}\xspace}

\def\Dz      {\ensuremath{\D^0}\xspace}
\def\Dzb     {\ensuremath{\Dbar^0}\xspace}
\def\DzDzb   {\ensuremath{\Dz {\kern -0.16em \Dzb}}\xspace}
\def\Dp      {\ensuremath{\D^+}\xspace}
\def\Dm      {\ensuremath{\D^-}\xspace}

\def\DpDm    {\ensuremath{\Dp {\kern -0.16em \Dm}}\xspace}
\def\Dstar   {\ensuremath{\D^*}\xspace}

\def\Dstarz  {\ensuremath{\D^{*0}}\xspace}

\def\B       {\ensuremath{\PB}\xspace}
  \def\Bbar    {\kern 0.18em\overline{\kern -0.18em \PB}{}\xspace}

\def\Bz      {\ensuremath{\B^0}\xspace}
\def\Bzb     {\ensuremath{\Bbar^0}\xspace}
\def\Bu      {\ensuremath{\B^+}\xspace}

\def\Bp      {\ensuremath{\Bu}\xspace}

\def\Bd      {\ensuremath{\B^0}\xspace}
\def\Bs      {\ensuremath{\B^0_s}\xspace}
\def\Bsb     {\ensuremath{\Bbar^0_s}\xspace}
\def\Bdb     {\ensuremath{\Bbar^0}\xspace}

  \def\Y#1S{\ensuremath{\PUpsilon{(#1S)}}\xspace}

\def\proton      {\ensuremath{\Pp}\xspace}

\def\BR         {{\ensuremath{\cal B}\xspace}}

\newcommand{\decay}[2]{\ensuremath{#1\!\to #2}\xspace}         

\def\to                 {\ensuremath{\rightarrow}\xspace}

\def\grpsuthree {\ensuremath{\mathrm{SU}(3)}\xspace}

\def\CP                {\ensuremath{C\!P}\xspace}

\def\AT#1     {\ensuremath{A_T^{#1}}\xspace}           

\def\C#1      {\ensuremath{\mathcal{C}_{#1}}\xspace}                       
\def\Cp#1     {\ensuremath{\mathcal{C}_{#1}^{'}}\xspace}                    
\def\Ceff#1   {\ensuremath{\mathcal{C}_{#1}^{\mathrm{(eff)}}}\xspace}        
\def\Cpeff#1  {\ensuremath{\mathcal{C}_{#1}^{'\mathrm{(eff)}}}\xspace}       
\def\Ope#1    {\ensuremath{\mathcal{O}_{#1}}\xspace}                       
\def\Opep#1   {\ensuremath{\mathcal{O}_{#1}^{'}}\xspace}                    

\newcommand{\tev}{\ensuremath{\mathrm{\,Te\kern -0.1em V}}\xspace}
\newcommand{\gev}{\ensuremath{\mathrm{\,Ge\kern -0.1em V}}\xspace}
\newcommand{\mev}{\ensuremath{\mathrm{\,Me\kern -0.1em V}}\xspace}
\newcommand{\kev}{\ensuremath{\mathrm{\,ke\kern -0.1em V}}\xspace}
\newcommand{\ev}{\ensuremath{\mathrm{\,e\kern -0.1em V}}\xspace}
\newcommand{\gevc}{\ensuremath{{\mathrm{\,Ge\kern -0.1em V\!/}c}}\xspace}
\newcommand{\mevc}{\ensuremath{{\mathrm{\,Me\kern -0.1em V\!/}c}}\xspace}
\newcommand{\gevcc}{\ensuremath{{\mathrm{\,Ge\kern -0.1em V\!/}c^2}}\xspace}
\newcommand{\gevgevcccc}{\ensuremath{{\mathrm{\,Ge\kern -0.1em V^2\!/}c^4}}\xspace}
\newcommand{\mevcc}{\ensuremath{{\mathrm{\,Me\kern -0.1em V\!/}c^2}}\xspace}

\def\cm   {\ensuremath{\rm \,cm}\xspace}

\def\mm   {\ensuremath{\rm \,mm}\xspace}

\def\nm   {\ensuremath{\rm \,nm}\xspace}

\def\barn{\ensuremath{\rm \,b}\xspace}

\def\invnb {\ensuremath{\mbox{\,nb}^{-1}}\xspace}

\def\invpb {\ensuremath{\mbox{\,pb}^{-1}}\xspace}

\def\invfb   {\ensuremath{\mbox{\,fb}^{-1}}\xspace}

\def\ns   {\ensuremath{{\rm \,ns}}\xspace}
\def\ps   {\ensuremath{{\rm \,ps}}\xspace}
\def\fs   {\ensuremath{\rm \,fs}\xspace}

\def\gsim{{~\raise.15em\hbox{$>$}\kern-.85em
          \lower.35em\hbox{$\sim$}~}\xspace}
\def\lsim{{~\raise.15em\hbox{$<$}\kern-.85em
          \lower.35em\hbox{$\sim$}~}\xspace}

\def\dllkpi     {\ensuremath{\mathrm{DLL_{K\pi}}}\xspace}

\def\mrad{\ensuremath{\rm \,mrad}\xspace}
\def\rad{\ensuremath{\rm \,rad}\xspace}

\def\gauss      {\mbox{\textsc{Gauss}}\xspace}

\def\tell1  {TELL1\xspace}
\def\ukl1   {UKL1\xspace}

\ifthenelse{\boolean{uprightparticles}}%
{\def\Prho      {\ensuremath{\uprho}\xspace}
 
}
{\def\Prho      {\ensuremath{\rho}\xspace}
 
}
\def\rhoz   {\ensuremath{\Prho^0}\xspace}

\def\Dpstarz  {\ensuremath{\D^{(*)0}}\xspace}

\def\Dsm  {\ensuremath{\D^{-}_{s}}\xspace}
\def\Dsp  {\ensuremath{\D^{+}_{s}}\xspace}

\def\DKstar    {\ensuremath{\Dz\Kstarz}\xspace}

\def\DbarRho   {\ensuremath{\Dzb\rhoz}\xspace}
\def\DRho   {\ensuremath{\Dz\rhoz}\xspace}

\def\pp    {\ensuremath{\proton\proton}\xspace}

\def\BtoDX   {\decay{\B}{\D \PX}}
\def\BtoDV   {\decay{\Bzb_{(d,s)}}{\Dz \PV}}
\def\BtoDKst {\decay{\Bdb}{\D \Kstarzb}}

\def\BstoDsPi   {\decay{\Bsb}{\Dsp \pim}}
\def\BdtoDmPi   {\decay{\Bdb}{\Dp \pim}}
\def\BptoDKp   {\decay{\Bp}{\D \Kp}}

\def\BsbartoDKstar      {\decay{\Bsb}{\Dz \Kstarz}}

\def\BsbartoDstarKstar  {\decay{\Bsb}{\Dstarz \Kstarz}}
\def\BsbartoDpstarKstar {\decay{\Bsb}{\Dpstarz \Kstarz}}
\def\BdbartoDKstar      {\decay{\Bdb}{\Dz \Kstarzb}}
\def\BdtoDKstar         {\decay{\Bdb}{\Dzb \Kstarzb}}
\def\BdtoDpipi         {\decay{\Bdb}{\Dz \pip \pim}}

\def\BdbartoDRho        {\decay{\Bdb}{\Dz \rhoz}}
\def\BdbartoDPiPi       {\decay{\Bdb}{\Dz \pipi}}

\def\KstartoKPi          {\decay{\Kstarz}{\Kp \pi-}}

\def\DtoKPiFav          {\decay{\Dz}{\Km \pip}}
\def\DtoKPiSup          {\decay{\Dz}{\Kp \pim}}
\def\DstartoDPi         {\decay{\Dstar}{\Dz\pion}}
\def\DsmtoKstarKm         {\decay{\Dsm}{\Kstarz \Km}}
\def\DmtoKstarKm         {\decay{\Dm}{\Kstarz \Km}}
\def\KstartoKPi         {\decay{\Kstarz}{\Kp \pim}}
\def\RhotoPiPi          {\decay{\rhoz}{\pip \pim}}
\def\btou    {\decay{\b}{\u}}
\def\btoc    {\decay{\b}{\c}}

\newcommand{\eq}[1]{Eq. \ref{equation : #1}}

\newcommand{\tab}[1]{Table \ref{table : #1}}
\newcommand{\fig}[1]{Fig. \ref{figure : #1}}

\newcommand{\ttst}[1]{\textrm{\scriptsize{#1}}}

\newcommand{\fitSigmaDRho}{\ensuremath{\sigma_{\ttst{\DbarRho}}\xspace}}
\newcommand{\fitSigmaDKstar}{\ensuremath{\sigma_{\ttst{\DKstar}}\xspace}}

\begin{document}

\begin{titlepage}
\pagenumbering{roman}

\vspace*{-1.5cm}
\centerline{\large EUROPEAN ORGANIZATION FOR NUCLEAR RESEARCH (CERN)}
\vspace*{1.5cm}
\hspace*{-0.5cm}
\begin{tabular*}{\linewidth}{lc@{\extracolsep{\fill}}r}
\ifthenelse{\boolean{pdflatex}}
{\vspace*{-2.7cm}\mbox{\!\!\!\includegraphics[width=.14\textwidth]{lhcb-logo.pdf}} & &}%
{\vspace*{-1.2cm}\mbox{\!\!\!\includegraphics[width=.12\textwidth]{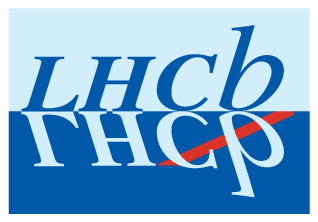}} & &}%
\\
 & & LHCb-PAPER-2011-008 \\
 & & CERN-PH-EP-2011-150 \\ 
 & & \today \\
 & & \\
\end{tabular*}

\vspace*{3.0cm}

{\bf\boldmath\huge
\begin{center}
First observation of the decay $\kern 0.18em\overline{\kern -0.18em B}{}^0_s \!\to D^0 K^{*0}$ and a measurement of the \vspace{0.2cm} ratio of branching fractions
$\frac{{\cal B}\left(\kern 0.18em\overline{\kern -0.18em B}{}^0_s \!\to D^0 K^{*0}\right)}{{\cal B}\left(\kern 0.18em\overline{\kern -0.18em B}{}^0 \!\to D^0 \rho^0\right)}$ 
\end{center}
}

\vspace*{0.2cm}
\begin{center}
{\small \it Submitted to Phys. Lett. B}
\end{center}

\vspace*{1.cm}
\begin{center}
The LHCb Collaboration
\footnote{Authors are listed on the following pages.}
\end{center}

\vspace{\fill}

\begin{abstract}
  \noindent
The first observation of the decay $\kern 0.18em\overline{\kern -0.18em B}{}^0_s \!\to D^0 K^{*0}$ using $pp$ data
collected by the LHCb detector at a centre-of-mass energy of 7 TeV, corresponding to an integrated luminosity of 36 pb$^{-1}$, is reported. 
A signal of $34.4 \pm 6.8$ events is obtained and the absence of signal is rejected with a statistical
  significance of more than nine standard deviations. The $\kern 0.18em\overline{\kern -0.18em B}{}^0_s \!\to D^0 K^{*0}$ branching fraction is measured 
  relative to that of $\kern 0.18em\overline{\kern -0.18em B}{}^0 \!\to D^0 \rho^0$: 
  $\frac{{\cal B}\left(\kern 0.18em\overline{\kern -0.18em B}{}^0_s \!\to D^0 K^{*0}\right)}{{\cal B}\left(\kern 0.18em\overline{\kern -0.18em B}{}^0 \!\to D^0 \rho^0\right)} = 
 1.48 \pm 0.34 \pm 0.15 \pm 0.12$, where the first uncertainty 
  is statistical, the second systematic and the third is due to the uncertainty on the ratio of the $B^0$ and $B^0_s$
  hadronisation fractions.
\end{abstract}

\vspace*{1.5cm}
\vspace{\fill}

\end{titlepage}

\newpage
\setcounter{page}{2}
\mbox{~}

\begin{center}
{\Large \bf The LHCb Collaboration}\\
\end{center}
\begin{flushleft}
{\small
R.~Aaij$^{23}$, 
C.~Abellan~Beteta$^{35,n}$, 
B.~Adeva$^{36}$, 
M.~Adinolfi$^{42}$, 
C.~Adrover$^{6}$, 
A.~Affolder$^{48}$, 
Z.~Ajaltouni$^{5}$, 
J.~Albrecht$^{37}$, 
F.~Alessio$^{37}$, 
M.~Alexander$^{47}$
% ...

{\it \footnotesize
$ ^{1}$Centro Brasileiro de Pesquisas F\'{i}sicas (CBPF), Rio de Janeiro, Brazil\\
% ...
$ ^{5}$Clermont Universit\'{e}, Universit\'{e} Blaise Pascal, CNRS/IN2P3, LPC, Clermont-Ferrand, France\\
$ ^{6}$CPPM, Aix-Marseille Universit\'{e}, CNRS/IN2P3, Marseille, France\\
% ...
$ ^{18}$Laboratori Nazionali dell'INFN di Frascati, Frascati, Italy\\
% ...
$ ^{23}$Nikhef National Institute for Subatomic Physics, Amsterdam, Netherlands\\
% ...
$ ^{35}$Universitat de Barcelona, Barcelona, Spain\\
$ ^{36}$Universidad de Santiago de Compostela, Santiago de Compostela, Spain\\
$ ^{37}$European Organization for Nuclear Research (CERN), Geneva, Switzerland\\
% ...
$ ^{42}$H.H. Wills Physics Laboratory, University of Bristol, Bristol, United Kingdom\\
% ...
$ ^{47}$School of Physics and Astronomy, University of Glasgow, Glasgow, United Kingdom\\
$ ^{48}$Oliver Lodge Laboratory, University of Liverpool, Liverpool, United Kingdom\\
% ...
$ ^{n}$LIFAELS, La Salle, Universitat Ramon Llull, Barcelona, Spain\\
% ...
}
}
\end{flushleft}

\cleardoublepage

\pagestyle{plain} 
\setcounter{page}{1}
\pagenumbering{arabic}


\section{Introduction}
A theoretically clean extraction of the Cabibbo-Kobayashi-Maskawa (CKM) unitarity triangle angle $\gamma$ can be performed using time-integrated \BtoDX decays by exploiting the interference 
between Cabibbo-suppressed \btou and Cabibbo-allowed \btoc transitions
\cite{Gronau1991172,Gronau1991483,Dunietz,PhysRevLett.78.3257,PhysRevD.63.036005,PhysRevD.68.054018}.
One of the most promising channels for this purpose is \BtoDKst, where \D represents a \Dz or a \Dzb meson.\footnote{In this Letter the mention of a decay will refer also to its charge-conjugate state.}
Although this channel involves the decay of a neutral \B meson, the final state is self-tagged by the flavour of the \Kstarz so that a time-dependent analysis is not required.
In the \BtoDKst decay, both the \BdbartoDKstar and the \BdtoDKstar are colour suppressed.
Therefore, although the \BtoDKst decay has a lower branching fraction compared to the \BptoDKp mode, it could exhibits an enhanced interference.

The Cabibbo-allowed \BsbartoDKstar and \BsbartoDstarKstar decays potentially provide a significant
background to the Cabibbo-suppressed \BdtoDKstar decay.
The expected size of this background is unknown, since
the \BsbartoDpstarKstar decay has not yet been observed.
In addition, a measurement of the branching fraction of \BsbartoDKstar is of
interest as a probe of \grpsuthree breaking in colour suppressed \BtoDV decays \cite{ColFerrandes,ChiangSenaha}, where \PV denotes a neutral vector meson.
Thus, the detailed study of \BsbartoDKstar is an important goal with the first LHCb data. 

The \lhcb detector~\cite{Alves:2008zz} is a forward spectrometer constructed to measure
decays of hadrons containing \b and \c quarks. 
The detector elements, placed along the collision axis of the Large Hadron Collider (\lhc),
start with the Vertex Locator, a silicon strip device that surrounds the \pp interaction region with its innermost sensitive part positioned \unit{8}{\mm} 
from the beam. 
It precisely determines the locations of the primary \pp interaction vertices, the locations of the decay vertices of 
long-lived hadrons, and contributes to the measurement of track momenta. Other tracking detectors include a large-area 
silicon strip detector located upstream of the \unit{4}{\Tm} dipole magnet and a combination of silicon strip detectors and straw drift chambers placed downstream. Two Ring-Imaging Cherenkov (\rich) detectors are used to identify charged hadrons. Further downstream an electromagnetic 
calorimeter is used for photon detection and electron identification, followed by a hadron calorimeter 
and a muon system consisting of alternating layers of iron and gaseous chambers. 
LHCb operates a two stage trigger system. In the first stage hardware trigger the rate is reduced from the visible interaction rate to about \unit{1}{\MHz} using information from the calorimeters and muon system.
In the second stage software trigger the rate is further reduced to \unit{2}{\kHz} by performing a set of channel specific selections based upon a full event reconstruction.
During the 2010 data taking period, several trigger configurations were used for both stages in order to cope with the varying beam conditions. 
\par
The results reported here uses \unit{36}{\invpb} of $\pp$ data collected at the LHC at a centre-of-mass energy
$\sqrt{s}=\unit{7}{\TeV}$ in 2010. The strategy of the analysis is to measure a ratio of branching fractions in
which most of the potentially large systematic uncertainties cancel. 
The decay \BdbartoDRho is used as the normalisation channel. 
In both decay channels, the \Dz is reconstructed in the Cabibbo-allowed decay mode
\DtoKPiFav; 
the contribution from the doubly Cabibbo-suppressed \DtoKPiSup decay is negligible.
The \Kstarz is reconstructed in the \KstartoKPi decay mode and the \rhoz in the \RhotoPiPi decay mode.
The main systematic uncertainties arise from the different particle
identification requirements and the pollution of the \BdbartoDRho peak by \BdbartoDPiPi decays where the \pipi pairs do not originate from a \rhoz resonance. 
In addition, the normalisation of the \Bs decay to a \Bd decay suffers
from a systematic uncertainty of \unit{8}{\percent} due to the current knowledge of the 
ratio of the fragmentation fractions
$f_s/f_d = 0.267^{ +0.021}_{-0.020}$ \cite{fsOverfd_LHCb}. 

\section{\texorpdfstring{Events selection}{Bd and Bs selection}}
Monte Carlo samples of signal and background events are
used to optimize the signal selection 
and to parametrize the probability density functions (PDFs) used in the fit. 
Proton beam collisions are generated with {\tt PYTHIA} \cite{pythia} and decays of hadronic particles are provided by {\tt EvtGen}~\cite{evtgen}. 
The generated particles are traced through the detector with {\tt GEANT4} \cite{geant},
taking into account the details of the geometry and material composition of the detector. 

\Bd and \Bs mesons are reconstructed from a selected \Dz meson combined with a vector particle (\rhoz or \Kstarz). 
The selection requirements are kept as similar as possible for \BsbartoDKstar and \BdbartoDRho. 
The four charged particles in the decay are each required to have a transverse momentum $p_T > \unit{300}{\MeVoverc}$
for the daughters of the vector particle and  $p_T > \unit{250}{\MeVoverc}$ (\unit{400}{\MeVoverc})
for the pion (kaon) from the \Dz meson decay.
The $\chi^2$ of the track impact parameter with respect to any primary 
vertex is required to be greater than 4.
A cut on the absolute value of the cosine of the helicity angle of the vector meson 
greater than 0.4 is applied.
The tracks of the \Dz meson daughters are combined to form a vertex with a goodness of fit $\chi^2/\textrm{ndf}$ smaller than 5. The \B meson vertex formed by the \Dz and the tracks of the \PV meson daughters is required to satisfy $\chi^2/\textrm{ndf}<4$.
The smallest impact parameter of the \B meson with respect to all the primary vertices is required to be smaller than 9 and defines uniquely the primary vertex associated to the \B meson.
Since the \Bd or \Bs should point towards the primary vertex, the angle between the \B momentum and the \B line of flight defined by the line between the \B vertex and the primary vertex is required to be less than $\unit{10}{\mrad}$. Finally, since the measured $z$ position (along the beam direction) of the \D vertex ($z_{\ttst{\D}}$) is not expected to be situated significantly upstream of the  $z$ position of the vector particle vertex ($z_{\PV}$), 
a requirement of $(z_{\ttst{\D}}-z_{\PV})/\sqrt{\sigma^2_{z\ttst{, \D}}+\sigma^2_{z\ttst{, \PV}}}>-2$ is applied, where $\sigma_{z\ttst{, \D}}$ and $\sigma_{z\ttst{, \PV}}$ are the uncertainties on the $z$ positions of the \D and \PV vertices respectively.

The selection criteria for the \PV candidates introduce
some differences between the signal and normalisation channel due to the particle identification (PID) and mass window
requirements.
The \Kstarz (\rhoz) reconstructed mass is required to be within \unit{50}{\MeVovercsq} (\unit{150}{\MeVovercsq}) of its nominal value \cite{pdg}. 
The selection criteria for the \Dz and vector mesons include identifying kaon and pion candidates using the RICH system. 
This analysis uses the comparison between the kaon and pion hypotheses, \dllkpi, which represents the difference in logarithms of likelihoods for the \kaon with respect to the \pion hypothesis.
The particle identification requirements for both kaon and pion hypotheses
have been optimized on data.
The thresholds are set at $\dllkpi>0$ and $\dllkpi<4$, respectively, for the kaon and the pion from the \Dz. The misidentification rate is kept low by setting the thresholds for the
vector meson daughters to $\dllkpi>3$ and $\dllkpi<3$ for the kaon and pion respectively. 
In order to remove the potential backgrounds due to \BstoDsPi and \BdtoDmPi with \DsmtoKstarKm and  
\DmtoKstarKm, 
vetoes around the nominal \Dm and \Dsm meson masses \cite{pdg} of $\pm \unit{15}{\MeVovercsq}$  are applied. Monte Carlo studies suggest that these vetoes are more than 99.5\% efficient on the signal.

Finally, multiple candidates in an event (about 5\%) are removed by choosing the \B candidate with the largest \B flight distance significance and which lies in the mass windows of the \Dz and the vector meson resonance.

\section{Extraction of the ratio of branching fractions}\label{section : extraction}
The ratio of branching fractions is calculated from the number of signal events 
in the two decay channels \BsbartoDKstar and \BdbartoDRho, 
\begin{eqnarray}\label{equation : ratio extraction}
\frac{\BR\left(\BsbartoDKstar\right)}{\BR\left(\BdbartoDRho\right)} = \frac{N^{\rm sig.}_{\ttst{\BsbartoDKstar}}}{N^{\rm sig.}_{\ttst{\BdbartoDRho}}}
\times
\frac{\BR\left(\RhotoPiPi\right)}{\BR\left(\KstartoKPi\right)}
\times
\frac{f_{d}}{f_{s}}
\times
\frac{\epsilon_{\ttst{\BdbartoDRho}}}{\epsilon_{\ttst{\BsbartoDKstar}}}
\end{eqnarray}
where the $\epsilon$ parameters represent the total efficiencies, 
including acceptance, trigger, reconstruction and selection, and
$f_s/f_d$ is the ratio of $B^0$ and $B^0_s$ hadronization fractions in \pp collisions at $\sqrt{s}=7$~TeV.
Since a given event can either be triggered by tracks from the signal or by tracks from the other {\it B} hadron decay, absolute efficiencies cannot be obtained with a great precision from the Monte Carlo simulation due to improper modelling of the generic \B hadron decays.
In order to reduce the systematic uncertainty related to the Monte Carlo simulation of the trigger,
the data sample is divided into two categories: candidates that satisfy only the hadronic hardware trigger\footnote{Events passing only the muon trigger on the signal candidate tracks are rejected.} 
($\texttt{TOSOnly}$, since they are Triggered On the Signal (TOS) exclusively and not on the rest of the event) 
and events which are Triggered by the rest of the event Independent of the Signal candidate \B decay ($\texttt{TIS}$). 
Approximately 6\% of candidates do not enter either of these two categories,
and are vetoed in the analysis.
The \BdbartoDRho signal yield is extracted separately for the two trigger categories $\texttt{TOSOnly}$ and $\texttt{TIS}$;
the \BsbartoDKstar signal yield is extracted from the sum of both data samples.
The ratio of efficiencies are sub-divided into the contributions arising from the selection
requirements (including acceptance effects, but excluding PID),  
$r_{\ttst{sel}}$, the PID requirements, $r_{\ttst{PID}}$,
and the trigger requirements, 
$r_{\ttst{\texttt{TOSOnly}}}$ and $r_{\ttst{\texttt{TIS}}}$.
The ratio of the branching fractions can therefore be expressed  as
\begin{eqnarray}\label{equation : masterEq}
\frac{\BR\left(\BsbartoDKstar\right)}{\BR\left(\BdbartoDRho\right)}  = 
\frac{\BR\left(\RhotoPiPi\right)}{\BR\left(\KstartoKPi\right)}
\times
\frac{f_{d}}{f_{s}}
\times
r_{\ttst{sel}}
\times
r_{\ttst{PID}}
\times
\frac{N^{\rm sig.}_{\ttst{\BsbartoDKstar}}}{\alpha\left(\frac{N^{\ttst{\texttt{TOSOnly}}}_{\ttst{\BdbartoDRho}}}{r_{\ttst{\texttt{TOSOnly}}}} + 
\frac{N^{\ttst{\texttt{TIS}}}_{\ttst{\BdbartoDRho}}}{r_{\ttst{\texttt{TIS}}}}\right)},
\end{eqnarray}
where $\alpha$ represents a correction factor for the ``non-\rhoz'' contribution in the \BdbartoDRho decays.

The values of the efficiency ratios are measured using simulated events, except for $r_{\ttst{PID}}=1.09 \pm 0.08$ which is obtained from data using the \DstartoDPi decay with \DtoKPiFav where clean samples of kaons and pions can be obtained using a purely kinematic selection.
Since the event selection is identical for the \Dz in the two channels of interest, 
many factors cancel out in $r_{\ttst{sel}}= 0.784 \pm 0.024$ thereby reducing the systematic uncertainties. 
The values of the trigger efficiency ratios, $r_{\ttst{\texttt{TOSOnly}}}= 1.20 \pm 0.08$ and $r_{\ttst{\texttt{TIS}}}= 1.03 \pm 0.03$,
depend on the trigger configurations and are therefore computed from a luminosity-weighted average. 
The quoted uncertainties reflect the difference between data and Monte Carlo simulation mainly caused by the energy calibration of the trigger.

The numbers of events in the two \DRho trigger categories, $N^{\ttst{\texttt{TOSOnly}}}_{\ttst{\BdbartoDRho}}$ and  
$N^{\ttst{\texttt{TIS}}}_{\ttst{\BdbartoDRho}}$, and $N^{\rm sig.}_{\ttst{\BsbartoDKstar}}$ are extracted from a simultaneous unbinned maximum likelihood fit to the data.
In order to simplify the description of the partially reconstructed background, the lower edge of
the \B meson mass window is restricted to \unit{5.1}{\GeVovercsq} for the \BdbartoDRho decay mode
and to \unit{5.19}{\GeVovercsq} for the \BsbartoDKstar decay mode.
There are four types of events in each category: signal, combinatorial background, partially reconstructed background and cross-feed.\footnote{The cross-feed events are due to particle misidentification on one of the vector daughters; some \DRho events can be selected as $\Dz\Kstarz$ and vice versa.}
The signal \B meson mass PDFs for \BdbartoDRho and \BsbartoDKstar are parametrized for each channel using
the sum of two Gaussians sharing the same 
mean value. 
The mean and width of the core Gaussian describing the \BdbartoDRho mass distribution are allowed to vary in the fit.
The fraction of events in the core Gaussian, $0.81 \pm 0.02$,
and the ratio of the tail and core Gaussian widths, $2.04 \pm 0.05$,
are fixed to the values obtained from Monte Carlo simulation.
In order to take into account the difference in mass resolution for the \BdbartoDRho and \BsbartoDKstar decay modes, 
the value of the ratio of core Gaussian widths $\frac{\fitSigmaDKstar}{\fitSigmaDRho} = 0.89 \pm 0.03$
is fixed from the Monte Carlo simulation.
The mass difference between the means of the \Bd and \Bs
signals is fixed to the nominal value~\cite{pdg}.

The combinatorial background mass distribution is modelled by a flat PDF and the partially reconstructed background is 
parametrized by an exponential function; the exponential slope is different in the \BdbartoDRho 
and \BsbartoDKstar categories.
Since the number of \BdbartoDRho decays is larger than that of \BsbartoDKstar, 
the contribution from misidentified pions as kaons from real \BdbartoDRho has to 
be taken into account. The fractions of the cross-feed events, $f_{\Dz \rhoz \to \Dz \Kstarz} = 0.062 \pm 0.031$ and $f_{\Dz \Kstarz \to \Dz \rhoz} = 0.095 \pm 0.047$, are constrained using the results from a Monte Carlo study corrected by the PID misidentication rates measured in data.
The PDF for the cross-feed is empirically
parametrised by a Crystal Ball function~\cite{CB2}, 
whose width and other parameters are taken 
from a fit to simulated events in which \BsbartoDKstar events are misidentified as \BdbartoDRho and vice versa;
the width is fixed to $1.75$ times the signal resolution.  
For the \BdbartoDRho decay mode, the events are further split according to the $\texttt{TOSOnly}$ and $\texttt{TIS}$ categories. 

In summary, 13 parameters are free in the fit.
Four shape parameters are used, two for the signal and two for the partially 
reconstructed backgrounds.
In addition, nine event yields are extracted, three (signal, combinatorial and partially reconstructed backgrounds) 
in each of the three categories: \BdbartoDRho ($\texttt{TOSOnly}$ and $\texttt{TIS}$) and \BsbartoDKstar. 

The results of the fit for \DRho and \BsbartoDKstar are shown in \fig{D0Rho_TOS_OtherB} and \fig{D0Kstar0_OS}. The overall signal yields are $154.1 \pm 15.1$ and $34.4 \pm 6.8$  respectively.
The yields for the different components are summarised in \tab{fit yield parameters}.

\begin{figure}[htbp]
\centering
\includegraphics[width=7.9cm]{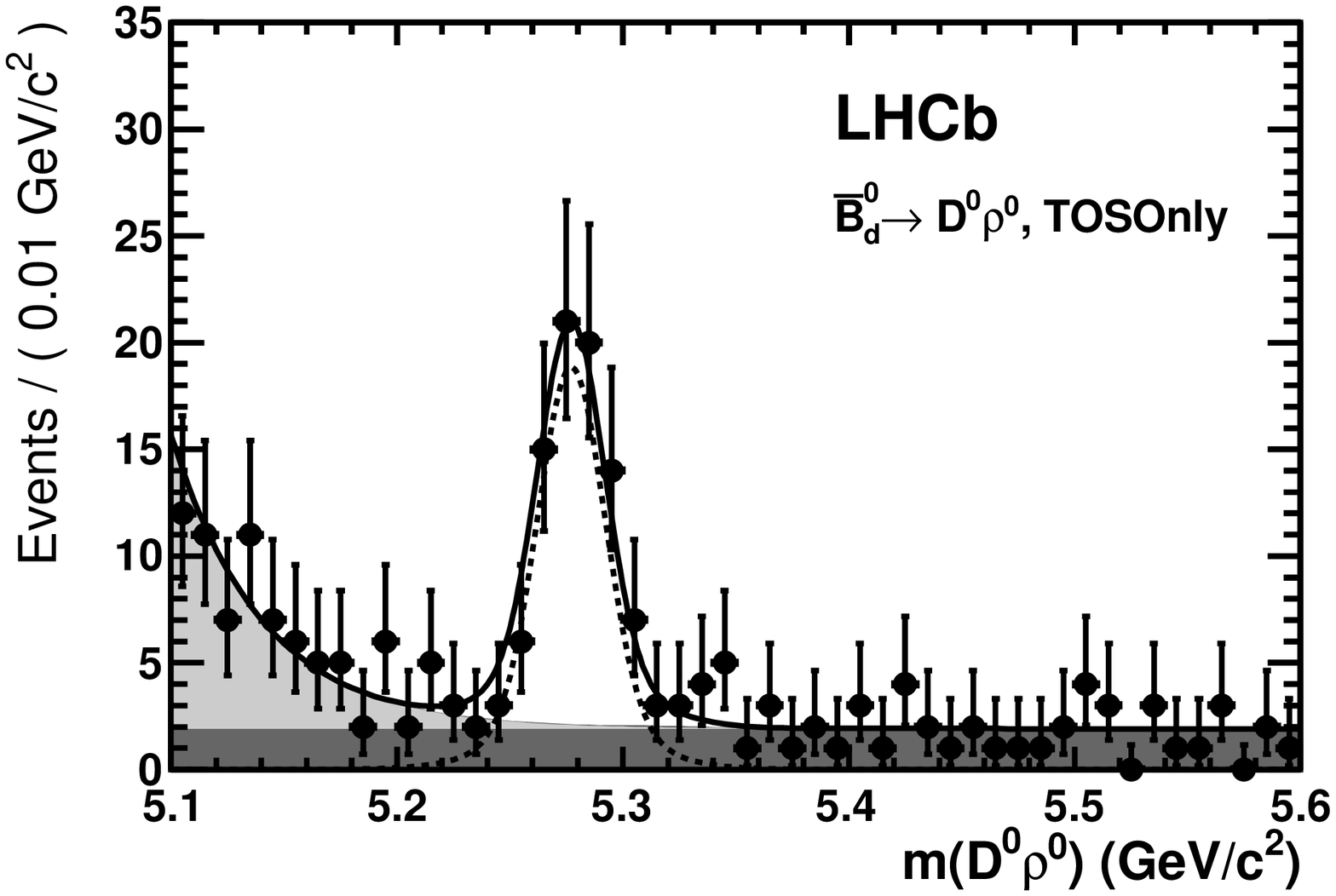}
\includegraphics[width=7.9cm]{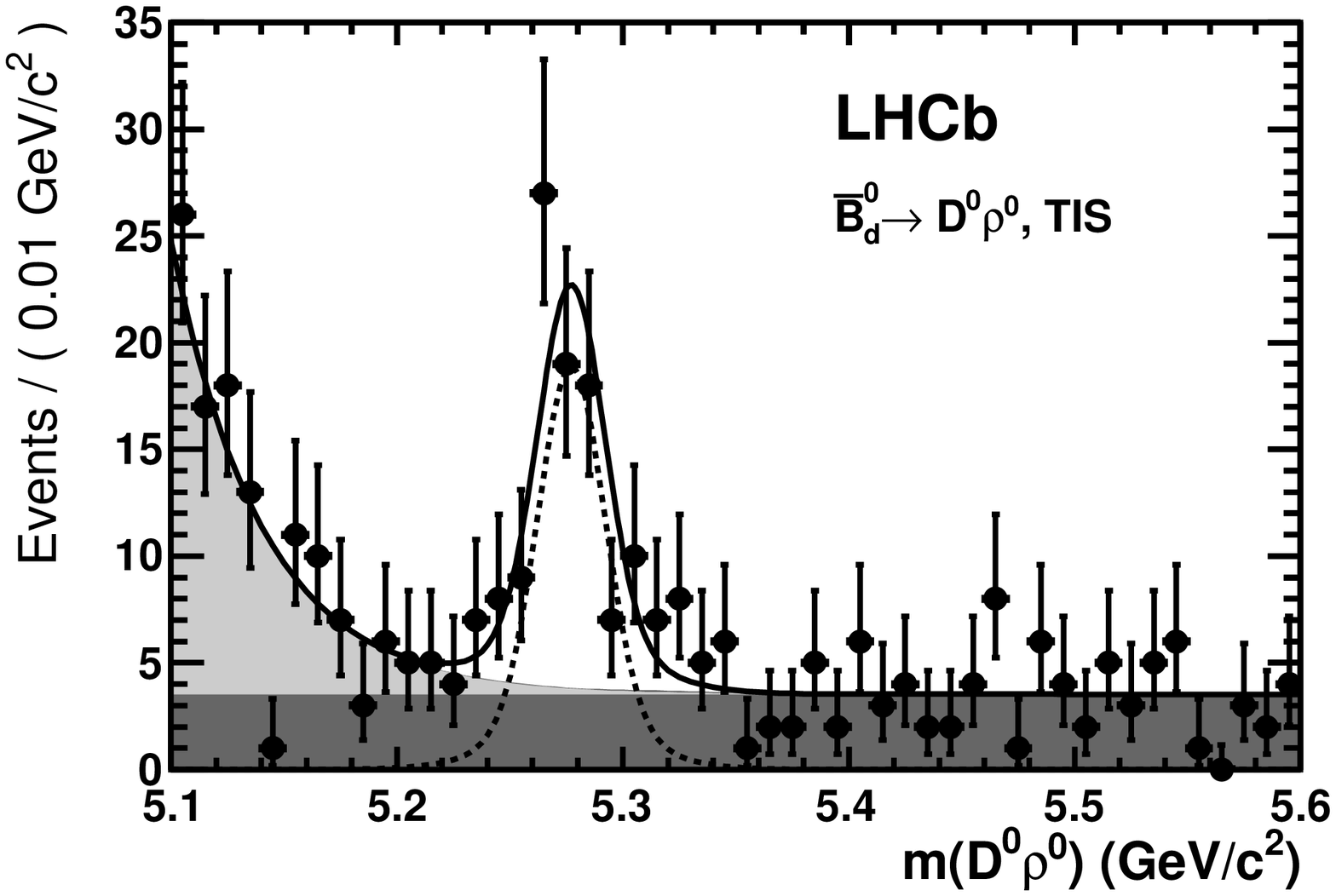}
\caption{The invariant mass distribution for the \BdbartoDRho decay mode for the $\texttt{TOSOnly}$ (left) and $\texttt{TIS}$ (right) trigger categories with the result of the fit superimposed. The black points correspond to the data and the fit result is represented as a 
solid line. The signal is fitted with a double Gaussian (dashed line), the partially reconstructed background with an 
exponential function (light grey area) and the combinatorial background with a flat distribution (dark grey area) 
as explained in the text. The contributions from cross-feed are too small to be visible.}
\label{figure : D0Rho_TOS_OtherB}
\end{figure}

\begin{figure}[htbp]
\centering
\includegraphics[width=12cm]{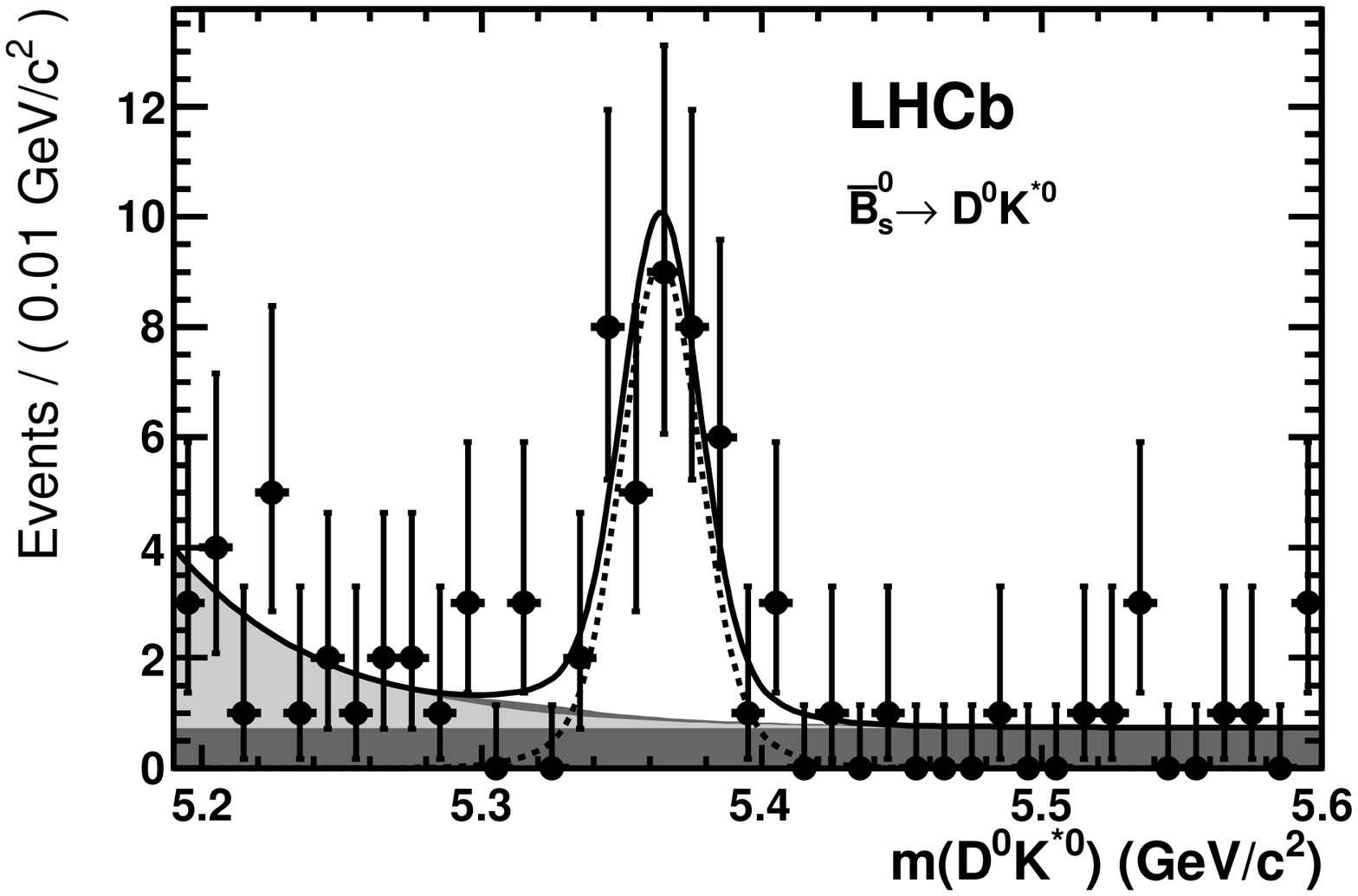}
\caption{The invariant mass distribution for the \BsbartoDKstar decay mode with the result of the fit superimposed. 
The black points correspond to the data and the fit result is represented as a 
solid line. The signal is fitted with a double Gaussian (dashed line), the partially reconstructed background with an 
exponential function (light grey area), the combinatorial background with a flat distribution (dark grey area) and the cross-feed from
 \BdbartoDRho (intermediate grey area) as explained in the text.}
\label{figure : D0Kstar0_OS}
\end{figure}

\renewcommand{\arraystretch}{1.}
\renewcommand{\arraystretch}{1.25}
\begin{table}[htb]
        \centering
        \caption{Summary of the fitted yields for the different categories. The background yields are quoted for the full mass regions.}
        \label{table : fit yield parameters}
        \begin{tabular}[h]{cccc}\hline
 Decay mode       &        Signal yield       & Part. rec. bkgd yield    &   Comb. bkgd yield \\\hline
\BsbartoDKstar                 & $34.4 \pm 6.8$  & $17.5 \pm 11.4$       & $29.8 \pm 8.4$ \\
\BdbartoDRho (\texttt{TOSOnly})& $77.0 \pm 10.1$  & $55.4 \pm 10.1$        & $95.5 \pm 13.1$ \\
\BdbartoDRho (\texttt{TIS}) & $77.1 \pm 11.2$ & $85.6 \pm 12.9$       & $176.0 \pm 17.5$ \\\hline
        \end{tabular}
\end{table}
\renewcommand{\arraystretch}{1.}

In order to check the existence of other contributions under the vector mass peaks, the sPlot technique~\cite{Pivk:2004ty} 
has been used to obtain background subtracted invariant mass distributions. The sWeights are calculated from the reconstructed \B invariant mass distribution using the same parametrization as in the analysis, the selection being the same except for the \PV invariant mass ranges which are widened. It was checked that there is no correlation between the \B and the \PV invariant mass. The resulting plots are shown in \fig{sPlots}, where the resonant component is fitted with a Breit-Wigner convoluted with a Gaussian and the non-resonant part with a second order polynomial. While the
\Kstarz region shows no sign of an extra contribution, the \rhoz region shows a more complicated
structure. An effective ``non-\rhoz'' contribution is estimated using a second-order polynomial: 
$30.1 \pm 7.9$ events contribute in the \rhoz mass window ($\pm \unit{150}{\MeVovercsq}$). The measured \BdbartoDRho yields are corrected by a factor $\alpha = 0.805 \pm 0.054$ (see \eq{masterEq}), consistent with expectations based on previous studies of the \BdtoDpipi Dalitz plot~\cite{Belle:Dalitz,Babar:Dalitz}. 

\begin{figure}[htbp]
    \centering
    \includegraphics[width=7.9cm]{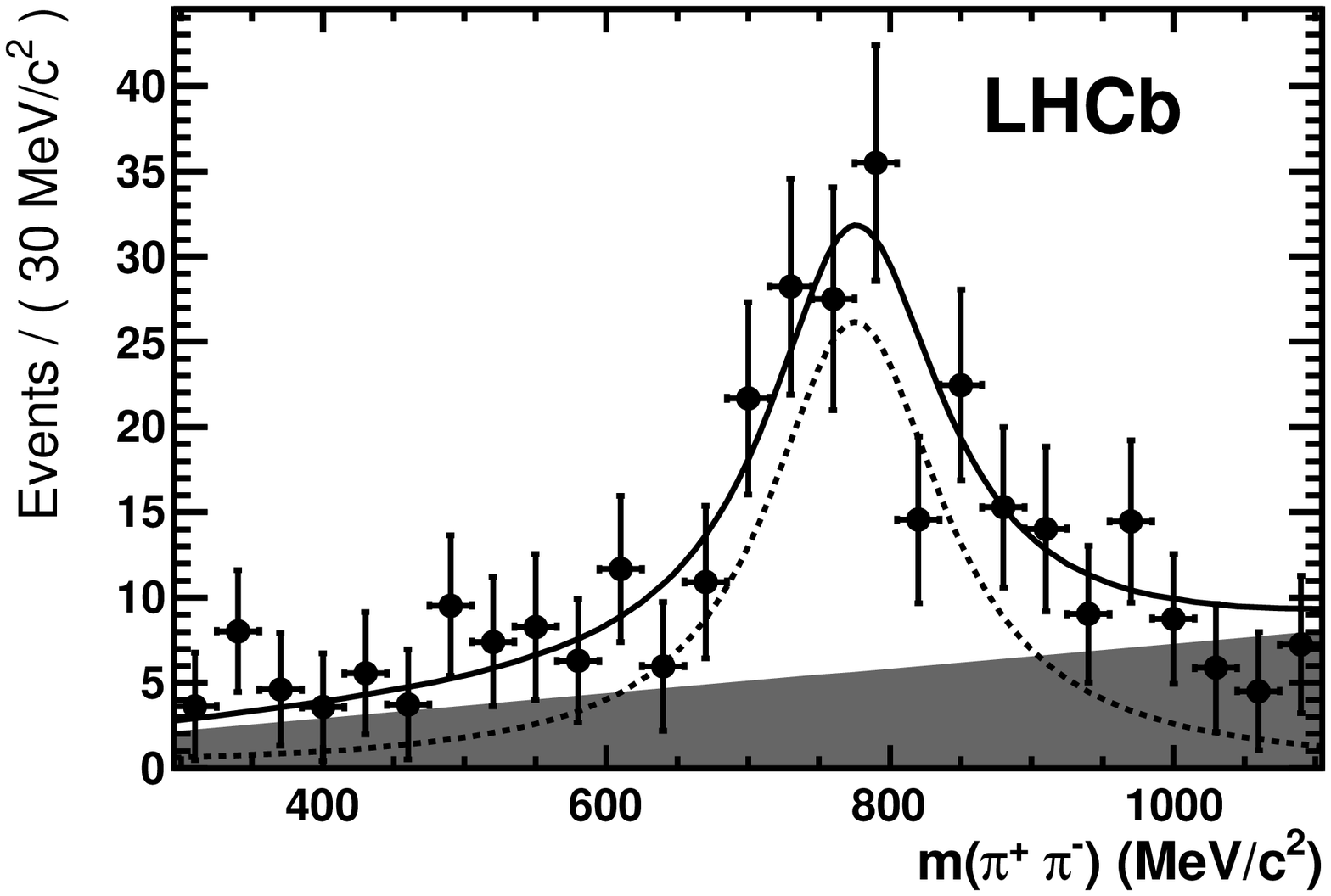}
    \includegraphics[width=7.9cm]{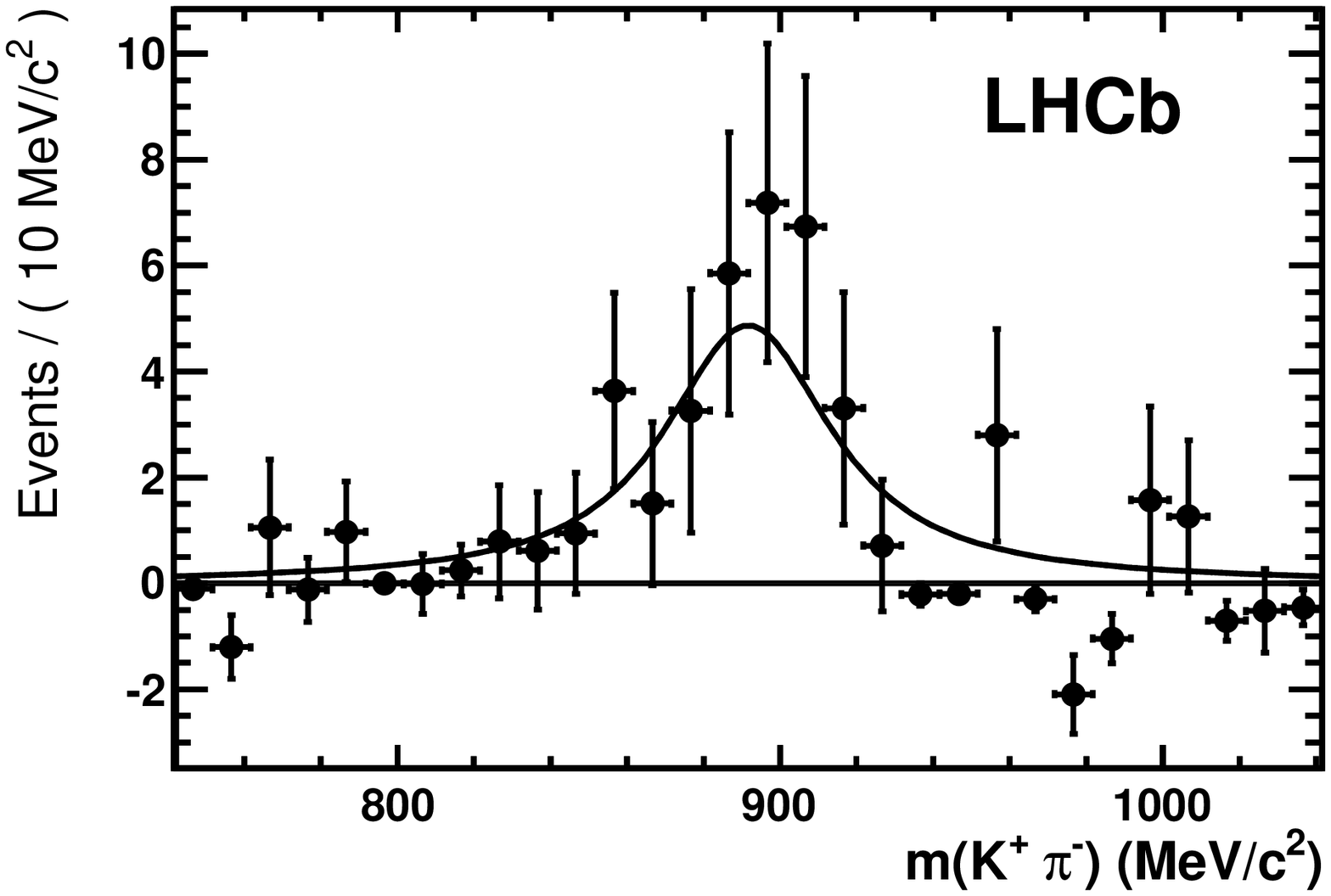}
    \caption{The \rhoz (on the left) and \Kstarz (on the right) invariant mass distributions obtained from data using an sPlot technique. The level of non \Kstarz combinations in the \BsbartoDKstar peak is negligible. 
    Despite being 
    mainly due to \DRho combinations, the \BdbartoDRho contains a significant contribution of ``non-\rhoz'' events. The black points correspond to the data and the fit result is represented as a solid line. 
    The resonant component is fitted with a Breit-Wigner convoluted with a Gaussian (dashed line) and the non-resonant part, 
    if present, with a second-order polynomial (grey area). }
     \label{figure : sPlots}
\end{figure}

The ratio of branching fractions,
$\frac{\BR\left(\BsbartoDKstar\right)}{\BR\left(\BdbartoDRho\right)}$,
is calculated using
the measured yields of the \BdbartoDRho signal in the two trigger categories, 
corrected for the ``non-\rhoz'' events and
assumed to contribute proportionally to the \texttt{TOSOnly} and \texttt{TIS} samples,
the \BsbartoDKstar yield 
and the values of the $r$ ratios quoted above.
The result is 
$\frac{\BR\left(\BsbartoDKstar\right)}{\BR\left(\BdbartoDRho\right)} =  1.48 \pm 0.34$, where the uncertainty is statistical only.
The small statistical correlation between the two yields due to the cross-feed has been neglected.

\section{Systematic uncertainties}
A summary of the contributions to the systematic uncertainty is given in \tab{systematics}.
The PID performances are determined with a \DstartoDPi data calibration sample reweighted
according to the kinematical properties of our signals obtained from Monte Carlo simulation. 
The systematic uncertainty has been assigned using the kinematical distributions 
directly obtained from the data. 
However, due to the 
small signal yield in the \Bs case, 
this systematic uncertainty suffers from large statistical fluctuations which directly translate into
a large systematic uncertainty on the kaon identification.
The statistical uncertainty obtained on the number of ``non-\rhoz'' events
present in the \rhoz the mass window ($\pm \unit{150}{\MeVovercsq}$) has been propagated in the systematic uncertainty. 
The differences observed between Monte Carlo simulation and data on the values of the \Dz and vector mesons reconstructed masses, as well as on the transverse momentum spectra, have been propagated into the uncertainty quoted on $r_{\ttst{sel}}$. 
The relative abundances of \texttt{TOSOnly} and \texttt{TIS} triggered events determined from simulated signal are in good agreement with those measured from data. 
This provides confidence in the description of the trigger in the
Monte Carlo simulation. 
Since these relative abundances are directly measured in data, they do not
enter the systematic uncertainty evaluation. 
However, the difference in trigger efficiency between the \BdbartoDRho and the
\BsbartoDKstar decay modes is taken from Monte Carlo simulation; this is
considered reliable since the difference arises due to the kinematical
properties of the decays which are well modelled in the simulation. 
The difference in the energy measurement between the hardware trigger
clustering and the offline reconstruction clustering is conservatively taken as a
systematic uncertainty due to the hadronic trigger threshold. 
The systematic uncertainty due to the $\texttt{TIS}$ trigger performances
on the two decay modes is obtained assuming that it does not depend on the
decay mode ($r_{\ttst{\texttt{TIS}}}=1$). 

The systematic uncertainty due to the PDF parametrizations has been evaluated using toy Monte Carlo simulations where the different types
of background have been generated using an alternative parametrization (wide Gaussians for the partially reconstructed backgrounds, first order polynomial for the 
combinatorial backgrounds) but fitted with the default PDFs.  

The total systematic uncertainty is obtained by combining all sources in quadrature.
The dominant sources of systematic uncertainty are of statistical nature and will be reduced with more data.
The error on the ratio of the fragmentation fractions \cite{fsOverfd_LHCb}
is quoted as a separate systematic uncertainty.
   
\begin{table}[htbp]
	\centering
		\caption{Summary of the contributions to the systematic uncertainties. The uncertainty on the $r$ ratio 
gives the range used for the systematic uncertainty extraction on the ratios of the branching fractions.}
	\label{table : systematics}
		\begin{tabular}[h]{lc}
			\hline
Source   & Relative uncertainty  \\\hline
Difference between data and MC to compute $r_{\ttst{PID}} = 1.09 \pm 0.06$  &  $\unit{5.8}{\percent}$\\
Uncertainty on the ``non-\rhoz'' component $\alpha = 0.805 \pm 0.054$ &   $\unit{6.8}{\percent}$\\ 
MC selection efficiencies $r_{\ttst{sel.}} = 0.784 \pm 0.024$  & $\unit{3.1}{\percent}$\\
\lone Hadron threshold $r_{\ttst{\texttt{TOSOnly}}} = 1.20 \pm 0.08$  &  $\unit{3.0}{\percent}$\\
\texttt{TIS} triggering efficiency $r_{\ttst{\texttt{TIS}}} = 1.03 \pm 0.03$  & $\unit{1.6}{\percent}$\\
PDF parametrisations  &   $\unit{1.0}{\percent}$ \\ \hline 
Overall relative systematic uncertainty &   $\unit{10.2}{\percent}$\\ \hline
Fragmentation fractions  &  $\unit{7.9}{\percent}$\\\hline
		\end{tabular}
\end{table}

\section{Summary}
A signal of $34.4 \pm 6.8$ \BsbartoDKstar events is observed for the first time.   
The significance of the background fluctuating to form the \Bs signal corresponds to 
approximately nine standard deviations,
as determined from the change in twice the natural logarithm of the likelihood of the fit without signal. Although this significance includes 
the statistical uncertainty only, 
the result is unchanged if the small sources of systematic error that affect the yields are included. 
The branching fraction for this decay is measured relative to that for
\BdbartoDRho, after correcting for the  ``non-\rhoz'' component, to be 
\begin{equation}
\frac{\BR\left(\BsbartoDKstar\right)}{\BR\left(\BdbartoDRho\right)} = 1.48 \pm 0.34 \pm 0.15 \pm 0.12,
\end{equation}
\noindent
where the first uncertainty 
is statistical, the second systematic and the third is due to the uncertainty in the hadronisation fraction ($f_s/f_d$).

The result is in agreement with other measurements of similar ratios and supports the \grpsuthree breaking observation in colour suppressed \BtoDV decays.
Using $\BR\left(\BdbartoDRho\right)= (3.2 \pm 0.5)\times 10^{-4}$ \cite{pdg} for the branching fraction of the normalising decay, a measurement of the 
\BsbartoDKstar branching fraction,
\begin{equation}
\BR\left(\BsbartoDKstar\right) = (4.72 \pm 1.07 \pm 0.48 \pm 0.37 \pm 0.74) \times 10^{-4},
\end{equation} 
is obtained, where the first uncertainty 
is statistical, the second systematic, the third due to the uncertainty in the hadronisation fraction ($f_s/f_d$) and the last 
is due to the uncertainty of the \BdbartoDRho branching fraction. 
A future, larger data sample will allow the use of the \BdbartoDKstar decay as the normalising channel, which will reduce the systematic uncertainty.

\section*{Acknowledgments}
We express our gratitude to our colleagues in the CERN accelerator
departments for the excellent performance of the LHC. We thank the
technical and administrative staff at CERN and at the LHCb institutes,
and acknowledge support from the National Agencies: CAPES, CNPq,
FAPERJ and FINEP (Brazil); CERN; NSFC (China); CNRS/IN2P3 (France);
BMBF, DFG, HGF and MPG (Germany); SFI (Ireland); INFN (Italy); FOM and
NWO (Netherlands); SCSR (Poland); ANCS (Romania); MinES of Russia and
Rosatom (Russia); MICINN, XuntaGal and GENCAT (Spain); SNSF and SER
(Switzerland); NAS Ukraine (Ukraine); STFC (United Kingdom); NSF
(USA). We also acknowledge the support received from the ERC under FP7
and the Region Auvergne.

\end{document}